**Title**: Human-Centered Development of an Explainable AI Framework for Real-Time Surgical Risk Surveillance

**Authors:** Andrea E. Davidson[1,2]*, Jessica M. Ray[1,3]*, Yulia Levites Strekalova[1,4], Parisa Rashidi[1,5], Azra Bihorac[1,2]

* These authors contributed equally to the present work

[1]Intelligent Clinical Care Center, University of Florida, Gainesville, Florida, USA

[2]Division of Nephrology, Hypertension, and Renal Transplantation, Department of Medicine, College of Medicine, University of Florida, Gainesville, Florida, USA

[3]Department of Health Outcomes and Biomedical Informatics, College of Medicine, University of Florida, Gainesville, Florida, USA

[4]Department of Health Services Research, Management and Policy, College of Public Health and Health Professions, University of Florida, Gainesville, Florida, USA

[5]J. Crayton Pruitt Family Department of Biomedical Engineering, University of Florida, Gainesville, Florida, USA

**Corresponding Author**

Azra Bihorac, MD, MS

Division of Nephrology, Hypertension, and Renal Transplantation, Department of Medicine, University of Florida, PO Box 100224, Gainesville, Florida, 32610-0224, USA

Phone: 1-352-294-8580

Email: abihorac@ufl.edu

## Abstract

**Background**: Artificial Intelligence (AI) clinical decision support (CDS) systems have the potential to augment surgical risk assessments, but successful adoption depends on their integration into clinical workflows, explainability, and user trust. This study reports the initial co-design of *MySurgeryRisk*, an AI CDS tool to predict the risk of nine post-operative complications in surgical patients.

**Methods**: Semi-structured focus groups and interviews were held as co-design sessions with perioperative physicians at a tertiary academic medical center in the Southeastern United States. Participants were first read a surgical vignette and were asked questions to elicit an understanding of their current workflow and decision-making practices before being introduced to the current *MySurgeryRisk* prototype web interface. Following the presentation of the prototype, participants were asked to provide feedback on the user interface and system features. Session transcripts were qualitatively coded, after which thematic analysis took place.

**Results**: Data saturation was reached after 20 surgeons and anesthesiologists from varying career stages participated across 11 co-design sessions. Thematic analysis of the focus group and individual interview transcripts and the coded segments resulted in five overall themes: (1) decision-making cognitive processes, (2) current approach to decision-making, (3) future approach to decision-making with *MySurgeryRisk*, (4) feedback on current *MySurgeryRisk* prototype, and (5) trustworthy considerations.

**Conclusion**: Clinical providers perceived *MySurgeryRisk* as a promising CDS tool that factors in a large volume of data and is computed in real-time without any need for manual input. Participants provided feedback on the design of the interface and imaged applications of the tool in the clinical workflow. However, its successful implementation will depend on its actionability and explainability of model outputs, integration into current electronic systems, and calibrate trust among end-users.

## Introduction

Surgical decision-making is a complex process that has been previously characterized as a cyclic model of ongoing information processing as unforeseen circumstances arise. Within this framework, surgeons continuously adapt their course of action through situational assessments, integrating new information, and decision adjustments [1]. Before a surgeon even steps foot in the operating room, decisions about surgical risk require the weighing of potential risks of surgery and post-operative complications with the benefits for a given patient. Such decisions occur along a spectrum of urgency (i.e., differentiating between urgent and emergent situations) and uncertainty [2]. The decision-making processes used within this context often involve judgment based on a clinician's prior experience and heuristics, particularly when decisions focus on complication risk [3]. In addition to this more naturalistic or intuitive decision-making process, studies suggest a second, more analytical decision process involving the intentional comparisons of potential courses of action, is also commonly applied in surgical settings [4]. Whether relying primarily on an analytical or intuitive approach, clinicians face the challenge of processing large amounts of data from an individual patient and comparing it to similar patient situations. With the increasingly large volume of available data in the clinical space, tools are needed to process and support the rapid processing of meaningful similarities between the patient at hand, similar patients, and cases with known outcomes.

Clinical decision support (CDS) systems offer an approach to overcoming these challenges. For example, the American College of Surgeons (ACS) National Surgical Quality Improvement Program (NSQIP) offers an online surgical risk calculator that uses 20 manually inputted patient characteristics to predict the likelihood of 18 unfavorable clinical outcomes occurring within 30 days of the operation [5]. In perioperative decision-making situations, predictive scores like NSQIP offer one method of quickly summarizing a patient compared to known data sets. However, when considering the cognitive needs of clinicians performing surgical risk predictions and the complexity of the data available, these traditional rule-based CDS systems may be of limited utility. The introduction of artificial intelligence (AI) algorithms to power CDS systems offers the promise of near real-time computation of complex patient data and comparisons made to similar patients from national and/or local data sets. Even so, incorporating the outputs of these more advanced algorithms into human decision-making can be limited by the human's ability to understand and interpret the information within the broader context of decision-making. This lack of understanding of the inner-working mechanisms of such systems is described as the "black box" problem in medical AI literature and has incited the growth of explainable AI research [6]. To successfully develop tools that can be quickly interpreted and incorporated into real-time analytical or intuitive decision processes, the design and presentation need to provide sufficient information for clinicians to incorporate model outputs through sensemaking into their existing mental models [7]. The problem of the "black box" can lead to distrust in the system. While less often considered within AI research, overtrust has been observed in similar automation settings such as auto-pilot and human-robot interactions [8], and may be anticipated within human-AI systems [9].

Explainable AI (XAI) is one approach to providing human users with information to calibrate their trust and incorporate the results of AI models into their decision-making. XAI is most commonly operationalized by presenting a set of features on which the AI prediction is built to explain how it reaches its predictions [10]. To develop meaningful XAI and understand how XAI will impact human decision-making in the clinical setting, it is crucial to involve the human end-user in the development of these tools. This engagement during development allows for understanding the user's current decision processes and identifies the necessary information to present in order to balance trust in the future XAI-powered CDS system [11].

User-centered design has been applied to both traditional CDS system development and paired development of CDS systems and implementation strategies [12]. This approach engages the end-user early and often to understand requirements and engage in iterative co-design of the tools and implementation strategies. The current work extends these methods to explore user requirements for a new XAI tool to support surgical risk decision-making. We report on the early stages of this iterative design process, where we have engaged users to understand current practices and tools used for surgical decision-making and to gather feedback on an XAI approach to support this type of decision-making.

## **Methods**

*Study Context*

This qualitative study applied a user-centered design to identify the information needs for perioperative risk decision-making and assessed formative usability of the *MySurgeryRisk* user interface, a prospectively validated machine-learning postoperative complication prediction tool developed with the longitudinal medical record data of over 60,000 patients who underwent an inpatient operation at our institution between 2014 and 2020. The *MySurgeryRisk* model computes risk scores for nine post-operative complications: mechanical ventilation >48 hours, intensive care unit (ICU) admission >48 hours, sepsis, acute kidney injury (AKI), neurological and delirium complications, venous thromboembolism, cardiovascular complications, surgical site healing complications, and 30-day hospital mortality.

Conducted as the first phase of an iterative lab-based co-design process involving perioperative providers in the design process, our objective was to identify the unique cognitive needs and preferences of clinical providers as end-users through semi-structured focus groups and interviews. The study setting was the main campus of a 1,111-bed tertiary care academic hospital with over 200 intensive-level care beds located in the Southeastern United States. This study received exempt Institutional Review Board approval.

*Participants*

We used purposive sampling to recruit staff at our institution from different roles in perioperative care and of varying stages in their career. Eligible participants were resident and attending physicians involved in perioperative care (i.e., surgeons, anesthesiologists, and intensivists). Potential participants were identified via the institution's directory and departmental listservs and were recruited via direct email solicitation. Written informed consent was obtained prior to participant enrollment in the study, and those who participated were compensated with a $50 Amazon gift card.

*Study Design*

Participants completed a demographic survey and participated in either an in-person focus group of 2-4 people or an individual semi-structured interview held over a Zoom call. The interviewers followed a semi-structured interview guide that was developed by a faculty member in our institution's Department of Health Outcomes and Biomedical Informatics, with clinical input from a surgical resident research fellow. This interview guide (Supplementary Materials 1) included a surgical patient vignette and probing questions related to participants' current clinical decision-making cognitive processes and surgical risk assessment approach. Once the participants' current workflow was construed, participants were given a demonstration of the *MySurgeryRisk* prototype web interface and its features, which displayed mock patient data, and were asked to provide feedback. Screenshots of the presented interface can be found in Supplementary Materials 2. Prominent interface prototype features included *MySurgeryRisk* model cards, patient characteristics and medical history, scheduled operation information, predicted risk scores for nine post-operative complications, and the complication outcomes of similar retrospective patients in the training cohort.

Qualitative sessions continued until data saturation was achieved. The sessions were recorded using Zoom, which generated a transcript of the session. A two-person approach was then used to reformat and edit the automated transcripts to produce a clean transcription using the session's audio recording. Filler words, stutters, and false starts were removed from transcripts.

*Analysis*

Qualitative interview and focus group transcripts were first analyzed using a rapid qualitative analysis approach to produce starting code categories. Two coders independently reviewed and coded all transcripts to identify current and future information needs and tool feedback. The coders met periodically throughout the process to discuss areas of discrepancy or uncertainty and iteratively revise the code categories and their definitions. At the conclusion of transcript coding, the coders met to identify the overarching themes and group the final code categories, presented in Table 2.

**Results**

*Co-Design Sessions and Participants*

Data saturation was achieved after 11 co-design sessions and a total of 20 participants. Of these 11 sessions, 5 were focus groups of 2-4 participants, and 6 were individual interviews. The participants included 5 acute care surgery attendings, 4 anesthesiology attendings, 7 general surgery residents, 3 neurosurgery residents, and 1 oral and maxillofacial surgical resident. A majority of participants (95%) were white, with a near-equal representation of men (55%) and women (45%) and a mean age of 40.2 years old. The complete participant demographics are presented in Table 1.

**Table 1**. **Demographic Characteristics of Participants**

| | *n* | % |
|---|---|---|
| **Gender** | | |
| Woman | 11 | 55 |
| Man | 9 | 45 |
| **Age Range** | | |
| 20-29 | 2 | 10 |
| 30-39 | 5 | 25 |
| 40-49 | 4 | 20 |
| 50-59 | 1 | 5 |
| 60-69 | 3 | 15 |
| **Race** | | |
| Asian | 1 | 5 |
| White | 19 | 95 |
| **Ethnicity** | | |
| Hispanic or Latino | 0 | 0 |
| Not Hispanic or Latino | 20 | 100 |
| **Level of training** | | |
| Attending | 9 | 45 |
| Resident | 11 | 55 |
| **Specialty** | | |
| Acute Care Surgery | 5 | 25 |
| General Surgery | 7 | 35 |
| Anesthesiology | 4 | 20 |
| Oral & Maxillofacial Surgery | 1 | 5 |
| Neurosurgery | 3 | 15 |
| **Years working at current organization** | | |
| 0-5 | 12 | 60 |
| 6-10 | 4 | 20 |
| 11-15 | 2 | 10 |
| 16-20 | 0 | 0 |
| 21-25 | 0 | 0 |
| 26-30 | 1 | 5 |
| 31-35 | 1 | 5 |

*Note.* *N* = 20. Participants were on average 40.2 years old (median: 34 years old, SD: 12.13).

*Qualitative Analysis*

Thematic analysis of the co-design session interview transcripts and the coded segments resulted in five overall themes: (1) decision-making cognitive processes, (2) current approach to decision-making, (3) future approach to decision-making with *MySurgeryRisk*, (4) feedback on current *MySurgeryRisk* prototype, and (5) trustworthy considerations. These themes and their associated coded segments are defined in Table 2, and representative quotes are presented in Table 3.

*Decision-Making Cognitive Processes*

This theme encompassed coded segments related to the timing of perioperative decisions, the information that providers use to assess surgical candidacy and risk, and the associated workflow challenges of gathering this information. When asked what data they rely on to inform surgical decision-making, participants consistently identified current medications, recent laboratory work, imaging, EKGs, patient comorbidities, and prior surgical history as essential. However, participants emphasized that these elements must be accompanied by sufficient clinical detail to understand how well controlled the comorbidities are. For instance, they noted that a label of "obesity" in the medical record is inadequate without a specific BMI, and similarly, cardiac or diabetic history needs to be contextualized with control measures such as blood pressure ranges or A1C levels. Providers described how previous surgical history and imaging findings guide the surgical approach. Scar tissue, for example, may influence the decision to perform open versus laparoscopic surgery. In patients with poorly controlled comorbidities or elevated operative risk, participants frequently discussed referring them for additional work-up or preoperative optimization, such as high-risk anesthesia consultations.

Beyond clinical data, participants emphasized the importance of understanding the patients' baseline functional status and goals of care. Functional status and exercise tolerance—like a patient's ability to walk a block or climb stairs—help to gauge physiological reserve and recovery potential. As one anesthesiology attending shared, "I'd want to know his functional capacity on the meds. If he's walking, running 10 miles a day, that's a whole different picture even if he looks [very sick] right in front of me rather than someone who's had trouble getting off the bed." Another anesthesiologist shared, "In general, if you can survive [a walk around the block], then you can probably survive the stress of most surgeries." Often times, such information can only be reliably obtained from the patient or their family themselves and providers explained that these clinic visits are critical to informing risk conversations and setting postoperative expectations. In some cases, such conversations influence a patient's decision to proceed with surgery. For instance, a patient or family might decline surgery upon understanding the likelihood of prolonged mechanical ventilation or discharge to rehabilitation facility.

The timing of perioperative decisions varied based on clinical context. For outpatient surgical planning, participants noted that risk assessment occurs weeks before the date of surgery, sometimes even before the clinic visit. Several reported making initial decisions about surgical candidacy while reviewing a patient's chart in advance of the visit, during which the conversation focuses on the logistics and expectations of surgery. As one surgeon noted, "that's when you're talking to the patient, you're really making your decision. That's when we would decide the type of surgery we're going to do and the route by which we're going to do it." Clinic visits—typically 2–4 weeks ahead of scheduled surgery—were identified as the window where shared decision-making crystallizes. In acute care settings, however, decision-making is more rapid, often occurring concurrently with diagnostic work-up, or in truly emergent cases, as the patient is being prepared for the OR.

Participants also described several challenges in their current cognitive workflows. The most frequently cited issue was the unreliability or completeness of the EMR data. Many noted that medication and problem lists are outdated, comorbidities are incompletely documented, and outside hospital information is often missing. One participant remarked, "Epic is filled with inconsistencies... if you're taking those inaccurate H&P findings and putting them into your prediction system, I'm not sure it's going to be accurate." To compensate, providers must often re-interview patients to verify histories and reconcile information across notes and systems. Another major limitation was the lack of structured or accessible data on functional status, frailty, cognitive impairment, and recent weight loss—factors that participants deemed highly influential but not reliably recorded. Some expressed a desire for these assessments to be standardized and integrated into the clinical record, but acknowledged this was often impractical due to time and resource constraints.

**Table 2. Identified Themes and Code Definitions**

| Theme | Code Category | Code Definition |
|---|---|---|
| Decision-Making Cognitive Processes | Information Informing Decision for Surgery/Surgery Risk | The information the provider is looking for and taking into consideration while assessing a patient's surgical risk or candidacy. |
| | Timing of Decision | Information regarding the time course of clinical decision-making, where decision support tools fit into workflow, and under what circumstances providers are willing to allot time to using a clinical decision tool. |
| | Existing Challenges | The existing challenges that participants identified in their workflow for gathering pertinent clinical information and making clinical decisions. |
| Current Approach to Decision Making | Decision Support | The providers' current use of surgical risk calculators, the standard operating procedures and guidelines currently used in decision-making, and opinions of currently used tools. |
| | Team Approach | The participation of other clinicians in surgical risk assessment and team involvement related to using current and future decision-support tools. |
| | Communication with Patients | The descriptions of how providers relay information to the patient and their families when presenting the option of surgical intervention, and how this clinical decision-support tool could aid this communication. |
| Future Approach to Decision Support with *MySurgeryRisk* | Decision Support | The providers' proposed future use of the *MySurgeryRisk* tool and its role in clinical decision-making. |
| | Team Approach | The imagined collaborative use of the *MySurgeryRisk* tool across provider specialties, roles, and perioperative disciplines. |
| | Communication with Patients | The descriptions of how providers imagine themselves using the *MySurgeryRisk* tool for communicating surgical risks with patients and their families. |
| Feedback on Current *MySurgeryRisk* Prototype | Prototype Feedback | Participant input on the prototype presented to them during their session and their thoughts and suggestions related to its usability and utility. |
| | Prototype Question | The questions participants had about the prototype that was presented to them and the features that were not clear to them. |

| | Type of Use | The different case uses for which participants could see themselves using the *MySurgeryRisk* decision-support tool. |
|---|---|---|
| Trustworthiness Considerations | AI Thoughts and Considerations | The participants' generalized beliefs, considerations, and warnings on artificial intelligence. Includes legal concerns, perceptions on the strengths and weaknesses of AI, implementation considerations, and what makes an AI system trustworthy. |
| | Prototype Challenges | The perceived challenges of the *MySurgeryRisk* modeling, data acquisition, and the elimination of human element to data synthesis, as they relate to surgical risk assessment. |

*Current Approach to Decision-Making*

This theme included coded segments related to participants' current use of decision-support tools, team-based approaches to surgical decision-making, and strategies for communicating surgical risk with patients. Across nearly all sessions, participants identified the American College of Surgeons National Surgical Quality Improvement Program (ACS NSQIP) risk calculator as the most commonly used clinical decision support tool for surgical risk stratification. Its widespread use was attributed not only to its availability but to the credibility of its published validation studies. As one participant remarked, "Like with NSQIP, there were a bunch of papers published on it, and you read the papers and they like fit, and you were like, 'Okay, fine, I'll use this black box.'" Other tools mentioned included the STOP-Bang questionnaire for sleep apnea risk and reference to published trial data and guidelines such as the Carotid Revascularization Endarterectomy Versus Stent (CREST) trials and the North American Symptomatic Carotid Endarterectomy Trial (NASCET) guidelines.

However, despite its use, participants often described the manual data entry into the NSQIP calculator as prohibitively burdensome, leading providers to only use it for patients they already consider as high-risk. The lack of data granularity to individual patients was identified as another limitation of NSQIP. As one surgical resident described, "It's a national database. Hundreds of thousands of patients and millions of deaths which have put in. But there's not enough granularity there to actually use that data for predicting the outcome of specific patients. It's kind of like quartiles and deciles and these sorts of things. But it but it's not really good for predicting risk in individual patients at all." Overall, NSQIP's outputs were often described as a resource to confirm one's suspicion of high risk, and as a tool to communicate this in a concrete numerical value, but may be often overridden by clinical judgment and experience, especially in complex or acute cases. Surgeons noted adjusting the predicted risk based on their own experience or the specifics of the surgical context, particularly when NSQIP lacked applicability, such as in emergent procedures.

Participants also discussed the collaborative nature of surgical decision-making, describing both informal and formal team-based practices. For some, the first introduction to a patient's case occurs through another provider's documented notes. Furthermore, one surgeon noted that although she doesn't calculate risk scores herself, she references the risk scores that her anesthesia colleagues typically include in their preoperative notes. Others shared that they typically defer more experienced colleagues to guide risk assessments. As one neurosurgery resident explained, "I rely on the experience of my attendings... they're like, 'Oh, it's a less than 1% risk,' and I'm [thinking], did you make that up?" This quote highlights the informal heuristics that shape surgical decision-making and the gaps that can emerge in knowledge transmission within the team from attending to resident.

The team approach expanded further when patients are deemed high-risk. In these situations, participants described involving other specialties—such as cardiology, pulmonology, or high-risk anesthesia—to help optimize patients prior to surgery. Consultations with the patient's other providers were commonly used to obtain surgical clearance or to coordinate medication adjustments. The multidisciplinary nature of this work was framed as essential to mitigate surgical risk and ensure patient safety.

Finally, participants emphasized that shared decision-making with patients is a central component of their current workflow because the decision to operate may ultimately lie with the patient. Elective surgical consultations were identified as a key moment not only for deciding whether to proceed with surgery, but also for setting expectations. Providers shared that communicating tangible risks—such as the likelihood of discharge to a nursing facility, prolonged intubation, or dialysis—was essential in helping patients and families understand the scope of the intervention and make informed decisions. These conversations were sometimes aided by decision-support tools, especially when numerical risk estimates were useful to underscore a message. As one participant put it, "These types of tools not only help decide whether to operate, but they help align expectations... you can say there's a 30% chance you won't go home." These discussions were seen not only as informative but also as decisive—prompting patients or families to reconsider whether surgery aligned with their values or goals of care.

*Future Approach to Decision Support with MySurgeryRisk*

This theme consisted of coded segments related to the providers' imagined future use of the *MySurgeryRisk* platform in their decision-making, team workflows, and communication with patients. Participants were generally optimistic about the tool's potential and discussed a variety of ways it could be integrated into their future practice. Many participants described using *MySurgeryRisk* during elective surgical planning, particularly in the outpatient clinic setting, where decisions about surgical candidacy and preoperative optimization typically occur. Participants emphasized that the platform could be most valuable during the initial stages of patient assessment—before the surgery is scheduled or the patient is booked for the OR. Some also pictured themselves checking *MySurgeryRisk* the morning or week of the operation to adjust their expectations and surgical plan. However, overall participants did not see this as an appropriate tool for deciding for or against surgery, "That feels icky to me... that has to be up to the provider," described one participant. In contrast to current tools like NSQIP, which require manual data entry, participants appreciated that *MySurgeryRisk* would automatically extract relevant information from the electronic health record in near real-time. As one provider noted, "This is something I could just have up on my desktop... it would auto-populate all this information." The reduced burden of manual input was seen as a key factor in promoting more consistent use of the tool.

Strong interest was expressed in *MySurgeryRisk* features that would allow them to explore modifiable risk factors. They described wanting to see how a patient's risk could change if certain preoperative conditions were improved, for example, better blood sugar control, smoking cessation, or medication adjustments. This would help guide shared decisions about whether to delay surgery and allow time for optimization. As one provider suggested, "If he quits smoking... what will happen?" These interactive or "what-if" features were envisioned as useful not only for clinical decision-making but also for patient education.

Team-based use of *MySurgeryRisk* also emerged as a common theme. Participants envisioned collaborating with colleagues in anesthesia, cardiology, and even primary care to use the platform as a shared reference point. The ability to incorporate the tool's output into multidisciplinary conversations, especially for high-risk patients, was seen as a way to improve alignment across specialties.

*MySurgeryRisk* was also seen as a valuable aid for patient communication. Participants anticipated using the tool to enhance informed consent discussions and to better convey individualized risk estimates. Risk outputs related to ICU admission, ventilator dependence, and discharge to a nursing facility were seen as particularly impactful. Visual displays or numeric summaries were described as helpful in "making it real" for patients and families, especially when preparing for complex or high-risk operations. One participant explained, "This could really help in terms of providing guidance," while another noted that seeing a concrete number or color-coded flag could influence how a patient or their caregiver perceives surgical risk.

*Feedback on Current MySurgeryRisk Prototype*

This theme included participant feedback and suggestions related to the *MySurgeryRisk* prototype shown to patients, including its user interface, functionality, and potential applications in clinical practice. Overall, participants responded positively to the prototype and expressed enthusiasm for the tool's future development. Several described the prototype as "impressive" or "exciting," noting that they would be interested in continued involvement as the tool evolves.

Participants shared a wide range of interface-specific feedback. Many emphasized the importance of a clean and intuitive design, noting that the current display was dense and could be overwhelming in a time-pressured clinical environment. Suggestions included organizing the layout of the pages more intuitively, clarifying the titles of each predicted complication, providing more information about risk score thresholds and interpretability, and being transparent about missing data. The visualization of risk scores—such as color-coded flags or complication icons—was seen as a strength of the current prototype. Participants felt that these visual elements could help quickly identify areas of concern and guide conversations with patients and families. Additionally, they appreciated the ability to "drill down" into each complication to see what data contributed to the prediction and asked for greater detail about the variables driving each risk. Different specialties also identified specific data elements they would want the model to incorporate—such as blood management and nutrition for anesthesia providers, or postoperative neurological changes for neurosurgery.

Nearly every participant emphasized the need for seamless integration into the current electronic medical record, Epic, stating that a separate platform would be burdensome to access. There was also widespread discussion on improvement to the model's actionability—participants emphasized that predictions and their identified features need to be those which can be addressed in the clinical setting. Without such actionability, several noted, the tool would just feel more like an alert system contributing to data fatigue rather than a useful tool they seek out to use on their own.

Participants offered several suggestions for welcomed features to the *MySurgeryRisk* platform. The most common suggestion was the ability to track the predicted versus actual complications over time for their patients as a form of performance feedback, displayed in a personalized dashboard view. Given that participants were eager to use this tool to support patient communication, some participants also proposed a simplified, patient-facing version of the interface for such conversations. Others requested addition was a prediction for discharge destination, such as likelihood of going home versus to a rehabilitation facility. This would allow providers and care coordinators to plan more proactively and help patients and families understand what to expect postoperatively.

Finally, the prototype prompted several outstanding questions about access, data management, and integration. Participants wanted to know who would have access to the tool, whether it would be embedded into existing EHR systems, and how the model would handle outdated or conflicting information.

*Trustworthiness Considerations*

This theme covered participant perspectives on the use of artificial intelligence in surgical decision-making, the considerations they felt needed to establish trust in a CDS tool like *MySurgeryRisk*, and the perceived challenges of the *MySurgeryRisk* system. Participants expressed a general openness to the use of AI in perioperative care, particularly if it could streamline workflows and support patient-provider conversations. However, trust in the system was described as highly contingent on transparency, accuracy, and alignment with clinical experience.

A major consideration for trust was the clarity and source of the data used by the model. Participants wanted to know which variables were included in the model, how much weight each carried in the final prediction, and what happened when data were outdated, missing, or conflicted with other parts of the chart. They emphasized the importance of transparency—both in terms of the data itself and the way predictions were calculated. As one provider explained, "What is the model based on? And where did the data come from?... I think those are common things that physicians are going to want to know." Providers noted that without this context, it would be difficult to trust or act on the tool's output, especially given that they already question the accuracy of certain data in the EMR.

Participants also raised concerns about the patient population the model was trained on. One focus group session pointed out that the system was trained on data from a single institution and was developed using retrospective data from patients who underwent surgery—omitting those whose cases were declined or delayed. This was seen as a fundamental limitation to its generalizability and ability to guide decisions about whether to operate. One provider explained, "if you're trying to look at the decision of whether or not to operate, those tools can't help you. Because they don't tell you what happened for all the patients who didn't have surgery. We're only looking at those who did."

Providers consistently stressed that while *MySurgeryRisk* could be a helpful input, it should not replace clinical judgment or the human elements of care. Many shared that their decisions are shaped by nuances that may not be captured in structured data, including frailty, patient goals, medication adherence, social circumstances, and recent changes in health status. They emphasized that trust in the tool would also depend on whether it could capture or contextualize this broader picture of the patient. Some also voiced concern about the ethical boundaries of algorithmic decision-making. For example, if the tool ever suggested whether to proceed with surgery, rather than simply offering a risk estimate, participants were clear that such recommendations would be inappropriate.

**Table 3. Representative Quotes Across Coded Categories**

| Theme | Code Category | Representative Quote |
|---|---|---|
| Decision-Making Cognitive Processes | Information Informing Decision for Surgery/Surgery Risk | "... understanding a surgical history could drastically change the type of operation that different surgeons would feel comfortable performing. [...] if they've had multiple operations, then you worry about the amount of scar tissue or adhesions that could make the complication rate higher" (surgery resident) |
| | Timing of Decision | "So as soon as a patient has an appointment registered you have to start building this model so the information's at hand during that initial assessment. Otherwise, you're having to call the patient back saying, 'I thought you were a good candidate for surgery, but I just saw this report and now you're not!'" (anesthesiology attending) |
| | Existing Challenges | "Because we find that unless you're really talking to someone, there are, I mean tons of things, and you can get a different answer from a patient talking to them five minutes from now. Medications and things that are not active or that they aren't taking like they're prescribed." (surgery resident) |
| Current Approach to Decision Making | Decision Support | "Yeah, so I'll use NSQIP calculator, but honestly I usually only use that in patients that I suspect will have higher-than-average risk. Just in your average patient getting an average surgery, so like a 30-something year old patient with maybe just obesity getting a cholecystectomy for symptomatic cholelithiasis, I wouldn't put that person in a risk calculator." (surgery resident) |
| | Team Approach | "I think we rely a lot on our anesthesia colleagues to say, 'Hey, this guy, I don't think we can get this guy through with their ejection fraction of XYZ,' you know what I mean? So, we rely a lot on them, and I look at their notes and see the scores they use" (surgery resident) |
| | Communication with Patients | "So, what I typically do is I will talk to the patient, and hopefully a family member at bedside, and tell them about both of the procedures that we do, one open and one closed. I will quote the data to them, I will tell them the risks and benefits of open, the risk and benefits of stent, and then give them my opinion on which one is appropriate for them based on their answers." (neurosurgery resident) |
| Future Approach to Decision Support with *MySurgeryRisk* | Decision Support | "Yeah that's awesome, it's a nice visual too. This could be linked in Epic and then when you open up their chart and you're sitting with them then you click on this link in Epic it pulls up their specific risk or something, and then you can show them how they can modify that risk—the things that are modifiable in a short period of time." (surgery resident) |

| | Team Approach | "And if we plan to be collaborative and involve primary care and have anesthesia input, which is absolutely essential, and again, I think the interfaces need to be somewhat personalized to the provider. So, for a primary care position, having this model provide post-discharge complication risks that can inform their frequency of follow up and the type of care that they provide a post-operative patient, and then intra-operative risk, if you think that's relevant for anesthesia care as well." (surgery resident) |
|---|---|---|
| | Communication with Patients | "So, I'm not sure that this would change my practice, but it could provide interesting information, or I guess maybe if a patient was on the fence about whether or not they wanted to have surgery this could be helpful with patient teaching that you can say, 'Yes, you have a heart condition, so you have X percent likelihood of getting of having a complication. But only 50% of people who are like you got them,' or something like that, like you can give them more numbers. Whereas right now I just say, 'Yeah, you're at higher risk than average.' And so, you could be a little more specific." (surgery resident) |
| Feedback on Current *MySurgeryRisk* Prototype | Prototype Feedback | "I would want to know what's been put into that risk to see if there are things that I can modify and delay the operation and improve that risk. So just knowing that it's higher and not understanding what it is, it doesn't help me." (surgery attending) |
| | Prototype Question | "Let's say, for example, it's trying to figure out like this risk factor here. But say the cholesterol labs are old, they're over a year old. Will it not be able to calculate it? Or will calculate it based on the old data?" (surgery attending) |
| | Type of Use | "So as far out as clinic and scheduling I would think this would be more useful. And then potentially morning of when the patient comes, if something happened you could pull this up and look at it as a pre-op evaluation on the morning of." (neurosurgery resident) |
| Trustworthiness Considerations | AI Thoughts and Considerations | "Yeah, a better understanding of the input variables would make it a little easier to trust right away. Otherwise, it's just going to be—like NSQIP there were a bunch of papers published on it, and you read the papers, and they like fit, and you were like okay, fine, I'll use this black box." (surgery attending) |
| | Prototype Challenges | "The histories that we encounter in Epic are frequently inaccurate. So, if you're taking those inaccurate H&P findings and then putting them into your prediction system, I'm not sure it's going to be accurate. I think that's basically the same thing you were saying." (surgery attending) |

## Discussion

This study describes findings from user centered design for the implementation of *MySurgeryRisk*, an AI-powered clinical decision support system designed to predict the likelihood of nine post-operative complications. Our findings provide insight into the types of information providers rely on for surgical decision-making and risk identification, their reactions to an early prototype of the *MySurgeryRisk* platform, and the key trust and usability factors that will determine future adoption.

Across focus groups and interviews, we identified varied decision-making processes driven by a combination of individual and team experience, available clinical data, patient goals, and team communication. While participants primarily rely on clinical gestalt and clinical guideline literature, participants reported limited use of predictive tools like NSQIP to confirm their assessment of a patient as being high-risk, and to support their communication of such risks more effectively to patients and other providers. Participants widely expressed interest in using *MySurgeryRisk*, welcoming it as an automated real-time system that eliminates the burden of manual data entry, identifies individualized modifiable risk factors, and provides interpretable risk scores. Given that participants reported using CDS tools primarily in situations where they already suspect high surgical risk, or referring to such scores in other providers' notes, our findings suggest that the autonomous *MySurgeryRisk* could enhance the current workflow by identifying unanticipated risks, empowering providers to take proactive action in response to important changes. In a previous study comparing the accuracy of provider risk assessments with those generated by *MySurgeryRisk*, the algorithm outperformed providers in predicting all complications except cardiovascular [13]. This study revealed that physicians' risk assessments were consistently lower than those of *MySurgeryRisk*, suggesting that risk may be currently under-estimated [13]. Similarly, a study by Sacks et al. also reported that surgeons underestimated the risk of surgical complications, and that exposure to risk calculator data resulted in more consistent and accurate risk assessments [5].

Our findings are consistent with prior work highlighting the importance of user trust, explainability, and alignment with clinical workflow in the successful implementation of AI systems in healthcare settings [10], [11]. Participants shared that a clear understanding of the input variables and their weights, data source, and descriptions of how each complication is defined, would increase trust in *MySurgeryRisk*. The availability of robust, peer-reviewed data on a CDS tool was consistently recognized as a key factor in fostering both its use and trust, highlighting that providers are strongly driven to practice evidence-based medicine. This understanding may present an important consideration for AI CDS system developers while publishing the results of such systems—that an effort should be made to write such manuscripts such that the language is accessible and clear to the clinical domain. From an interface standpoint, participants valued the potential to drill down into contributing variables, visualize predicted complications, and receive actionable guidance for early interventions and preoperative optimization. The interface received design recommendations to improve usability, including simplified layouts and clearer titles, integration into the current electronic medical records (EMR), and the addition of outcomes such as discharge destination or functional capacity.

Our research identified key user concerns that could impact future adoption. Namely, data quality was a consistent point of discussion. As is common with AI tools, our prediction model is based on structured retrospective data from the institution's electronic health record. While multiple systems are in place to ensure data integrity, clinicians noted that data in the EHR are often inconsistent—whether from inconsistencies in patient-provided histories, data entry, or missing information from outside institutions. Thus, efforts to mitigate such inconsistencies by the model will need to be made transparent to clinician users to mitigate their impact on user trust in the model.

The results presented in this study draw from the diverse inclusion of participants across levels of training, perioperative specialties, gender, and age. Thus, we have captured a representative range of perspectives from early- and late-career physicians. Our co-design methodology is strengthened by the inclusion of providers from the institution where the *MySurgeryRisk* training data was sourced, to further the relevance of the findings. By engaging directly with this population, we ensure that the insights gathered are aligned with real-world experience with these patients and the specific needs and challenges faced by clinicians at various stages of their careers. However, this focus also presents a limitation. The findings may not be generalizable to institutions with differing patient populations, practices, or attitudes toward AI-based tools. Participants in this study were also largely white and drawn from an academic institution actively involved in AI research, which may limit the transferability of results to other, community-based settings. Finally, while this study captured the views of physicians, future work is needed to incorporate the perspectives of other stakeholders such as nurses, patients, and hospital management. In the next stage of co-design, providers will take part in "think aloud" sessions to further test the updated interface's usability prior to an implementation period.

## Conclusion

This study reports the results of the initial co-design sessions of *MySurgeryRisk*, an explainable AI CDS tool for surgical risk prediction. Participants identified a need for tools that integrate seamlessly into clinical workflows, provide modifiable and actionable insights, and offer transparency in how predictions are generated. These findings will inform the continued development and refinement of *MySurgeryRisk*, and more broadly, can guide best practices in the design, implementation, and trust of AI CDS tools in perioperative care. Future research will further test the usability of the user interface and will investigate the perspectives of other perioperative care stakeholders.

**<u>Supplementary Materials</u>**

Human Centered Development of an Explainable AI Framework for Real-Time Surgical Risk
Surveillance

**Supplementary Materials 1. Co-Design Session Semi-Structured Interview Guide**

Format: Small Groups (3-5 participants) or Individual

Goals: Identify user cognitive needs and context of use

NOTE: Need to reinforce physiological safety

Setup: Recording equipment; Index Cards; Design Poster Sheets

*Thank you for joining us today. We have asked you here today to assist us with the initial design concepts of an interactive user interface that will support clinician interaction with an AI system designed to aide with surgical risk decision making. Our goal is to develop an interface that is easy to use in the context of decision making and that will provide clinicians the necessary information to easily understand and interact with the information generated by the AI models.*

*We will start this session with understanding how you currently assess surgery risk*... I am going to read a patient story to you and I want the group to discuss what you, in your current practice, would decide for this patient. We know that there are often many paths to clinical decisions and that individual clinicians may process different information, so please be open and honest and remember that no answers should leave this group.

Your first patient, Mr. Jones is a 67-year-old male admitted to undergo laparoscopic distal pancreatectomy for a main duct–intraductal papillary mucinous neoplasm (MD-IPMN) in the distal third of pancreas. Patient Medical History: hyperlipidemia, hypertension (HTN), Type II diabetes (DMII), Obesity. Patient reports being a half pack a day smoker for 30 years. Current BMI 32.

1. "What additional information would you expect to have about this patient at this stage?" *(let them fill in the blanks with patient details)*

2. "What additional information would you not necessarily have, but would like to have at this stage?"

3. *Probe each member of the group*...What would be your next course of action?"...

4. *Once each member has spoken*..."Why did you choose this course of action? What information stood out to you as important? Why was it important?" *Repeat for each member, emphasizing differences in different decisions.*

Now that we have uncovered some differences in decision processes, I want to move to the prototype...*we are going to demonstrate our current tool design for you. Please feel free to ask questions and we will also provide prompts for feedback as we go through the demonstration.*

**Supplementary Materials 2. *MySurgeryRisk* Prototype Web Interface Screenshots**

Note: Data presented in the below screenshots are not real patients' data.

# MySurgeryRisk

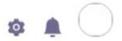

Cases

| TODAY | TOMORROW | |

🔍 Search Users by Name , ID, Room Number

All | Scheduled | In Progress | Completed | Cancelled

| MRN | Name | Scheduled Time | Procedure | Complications | |
|---|---|---|---|---|---|
| dsggglh7 | John<br>Age: 35   Male | ● Completed<br>19/Sept/2023 20:00 | Heart<br>Room: HBI 12 | | ☆ |
| dsggglh8 | Emma<br>Age: 28   Female | ● Completed<br>7/Nov/2023 20:00 | Knee<br>Room: HBI 19 | | ⭐ |
| dsggglh9 | Michael<br>Age: 45   Male | ● In Progress<br>17/Oct/2023 20:00 | Abdominal<br>Room: HBI 16 | | ⭐ |
| dsggglh10 | Sophia<br>Age: 32   Female | ● Completed<br>29/Sept/2023 20:00 | Spine<br>Room: HBI 8 | | ☆ |
| dsggglh11 | William<br>Age: 38   Male | ● In Progress<br>9/Oct/2023 20:00 | Lung<br>Room: HBI 14 | | ☆ |
| dsggglh12 | Olivia<br>Age: 27   Female | ● Completed<br>14/Nov/2023 19:00 | Eye<br>Room: HBI 21 | | ☆ |
| dsggglh14 | Emily<br>Age: 27   Female | ● Completed<br>13/Oct/2023 20:00 | Eye<br>Room: HBI 11 | | ☆ |
| dsgglhd | Benjamin<br>Age: 33   Male | ● In Progress<br>14/Oct/2023 20:00 | Lung<br>Room: HBI 22 | | ☆ |

| | Daniel<br>Age: 37    Male | ● Cancelled<br>16/Oct/2023 20:00 | Shoulder<br>Room: HBI 9 | ☆ |
|---|---|---|---|---|
| dsrggjh17 | | | | |
| dsrggjh18 | Charlotte<br>Age: 31    Female | ● In Progress<br>17/Oct/2023 20:00 | Kidney<br>Room: HBI 12 | ☆ |

PREV    NEXT

## Model Cards

Overall    AKI    ICU/Admission    Mechanical Ventilation    Sepsis    Wound    Neuro and Delrism    Cardiovascular    Venous Thromboembolism    Mortality

The AI analytics system analyzed a dataset of 10,000 patients.

**Overall, 30% patients would get complications.**

**Common Medical Conditions:** The most prevalent medical conditions among the analyzed patients were hypertension (45%), diabetes (30%), and asthma (15%).

**Age Distribution:** The majority of patients with complications fell within the age range of 50-65, comprising 40% of the analyzed population.

**Comorbidity Trends:** The system identified a high incidence of comorbidities among patients with complications. For instance, 60% of patients with diabetes also had hypertension.

### Complications Assessment

**Complication Frequency:** Among the analyzed patient population, 20% experienced complications related to their medical conditions.

**Most Common Complications:** The most frequently observed complications were cardiovascular events (10%), kidney dysfunction (8%), and respiratory infections (5%).

**Risk Factors:** The AI analytics system identified specific risk factors associated with each complication, such as smoking habits, obesity, and medication adherence.

### Recommendations and Interventions

**Personalized Recommendations:** Based on the AI analytics system's analysis, it provided personalized recommendations for patients to mitigate the risk of complications. For instance, patients with hypertension were advised to maintain a healthy diet, exercise regularly, and adhere to prescribed medication.

**Intervention Strategies:** The system suggested proactive interventions such as regular health screenings, lifestyle modifications, and medication adjustments for patients at higher risk of complications.

# Patient's Complications

## John 👥

### Basic Info
Age: 35
Gender: Male
Admitted: 2023-10-20 20:30

Insurance: 300fid
Residence: Newberry 203, FL

### Scheduled Surgery:
Room: HBI 12
Date: Tue Sep 19 2023
Type: Heart

### Comorbities
The patient has a history of 2 complications Venous Thromboembolism, Mechanical Ventilation > 48Hrs

### Medications
The patient has taken 0 medications within the past year

## Procedures
The patient was admitted to the hospital today on a weekday from a non-transfer admission setting, and is scheduled to have an elective plastic surgery, specifically an unspecified primary procedure.

## Labs
Within the one week of surgery, there were no blood HGB tests and no urine HGB tests, and the average laboratory values are as follows: HGB not available, serum platelets not available.

### Risks

| | | |
|---|---|---|
| **Acute Kidney Injury** score: 89% Updated: Sep 25, 2022, 13:25 PM ⊗ ✓ | **ICU Admission > 48Hrs** score: 89% Updated: Sep 25, 2022, 13:25 PM ⊗ ✓ | **Mechanical Ventilation > 48Hrs** score: 89% Updated: Sep 25, 2022, 13:25 PM ⊗ ✓ |
| **Wound Complication** score: 89% Updated: Sep 25, 2022, 13:25 PM ⊗ ✓ | **Neuro Complication and Delirium** score: 89% Updated: Sep 25, 2022, 13:25 PM ⊗ ✓ | **Cardiovascular Complication** score: 89% Updated: Sep 25, 2022, 13:25 PM ⊗ ✓ |
| **Sepsis** score: 89% Updated: Sep 25, 2022, 13:25 PM | **Venous Thromboembolism** score: 91% Updated: Sep 25, 2022, 13:25 PM | **30 Days Hospital Mortality** score: 89% Updated: Sep 25, 2022, 13:25 PM |

2023-11-27 17:23:14

**Sepsis**
Updated: Sep 25, 2022, 13:25 PM
score: 89%

**Venous Thromboembolism**
Updated: Sep 25, 2022, 13:25 PM
score: 91%

**30 Days Hospital Mortality**
Updated: Sep 25, 2022, 13:25 PM
score: 89%

## Medical History

**Labs**  Procedures  Medications  Vitals

### Lastest Blood Test

| Test | Value | Time | Status |
|---|---|---|---|
| Blood Pressure | 120/80 mmHg | Less than 130/80 mmHg | 2023-06-28 09:30 |
| Total Cholesterol | 180 mg/dL | Less than 200 mg/dL | 2023-06-27 14:45 |
| HDL Cholesterol | 50 mg/dL | Greater than 40 mg/dL | 2023-06-26 11:20 |
| LDL Cholesterol | 110 mg/dL | Less than 130 mg/dL | 2023-06-25 17:10 |
| Triglycerides | 120 mg/dL | Less than 150 mg/dL | 2023-06-24 10:55 |
| Hemoglobin A1c | 6.0% | Less than 7.0% | 2023-06-23 08:15 |
| Fasting Blood Glucose | 100 mg/dL | Less than 100 mg/dL | 2023-06-22 12:30 |

### Lastest Urine Test

| Test | Value | Time | Status |
|---|---|---|---|
| Urinalysis | Normal | - | 2023-06-28 09:30 |
| pH | 6.5 | 5.0-7.0 | 2023-06-27 14:45 |
| Specific Gravity | 1.020 | 1.005-1.030 | 2023-06-26 11:20 |
| Protein | Negative | - | 2023-06-25 17:10 |
| Glucose | Negative | - | 2023-06-24 10:55 |
| Ketones | Negative | - | 2023-06-23 08:15 |
| Blood | Negative | - | 2023-06-22 12:30 |

## Similar Patient Outcomes

**Acute Kidney Injury**
**Score:**  Avg: 5%  Patient: 3.3%
12 cases  Similarity: **80%**

**Cardiovascular**
**Score:**  Avg: 5%  Patient: 3.3%
12 cases  Similarity: **90%**

**Sepsis**
**Score:**  Avg: 5%  Patient: 3.3%
12 cases  Similarity: **10%**

**ICU Admission > 48Hrs**
**Score:**  Avg: 5%  Patient: 3.3%
12 cases  Similarity: **72%**